\documentclass[fleqn,usenatbib]{mnras}

 \usepackage{amsmath}
\usepackage{txfonts}

\usepackage[T1]{fontenc}

\DeclareRobustCommand{\VAN}[3]{#2}
\let\VANthebibliography\thebibliography
\def\thebibliography{\DeclareRobustCommand{\VAN}[3]{##3}\VANthebibliography}

\usepackage{graphicx}	
\usepackage{amsmath}	

\usepackage{hyperref}
\usepackage{rotating}
\usepackage{longtable}
\usepackage{multirow}
\usepackage[normalem]{ulem}
\usepackage[flushleft]{threeparttable}
\usepackage{bm}
\usepackage{makecell}
\usepackage{array}

\usepackage{gensymb}
\usepackage{ragged2e}

\usepackage{xcolor}

\newcommand{\hii}{H\textsc{ii}}
\newcommand{\hb}{\mathrm{H}\beta}
\newcommand{\ha}{\mathrm{H}\alpha}
\newcommand{\Oiii}{\mathrm{[O\ III]}}

\title[The \hii\ Galaxy Hubble Diagram]
{Model selection using the \hii\ galaxy Hubble diagram}

\author[Wei \& Melia]
{Jun-Jie Wei$^{1,2}$\thanks{E-mail: jjwei@pmo.ac.cn}
and Fulvio Melia$^{3}$\thanks{John Woodruff Simpson Fellow. Email: fmelia@email.arizona.edu}
\\
$^{1}$Purple Mountain Observatory, Chinese Academy of Sciences, Nanjing 210023, China\\
$^{2}$School of Astronomy and Space Sciences, University of Science and Technology of China, Hefei 230026, China\\
$^{3}$Department of Physics, The Applied Math Program, and Department of Astronomy, The University of Arizona, AZ 85721, USA
}

\date{Accepted XXX. Received YYY; in original form ZZZ}

\pubyear{\the\year{}}

\begin{document}

\label{firstpage}
\pagerange{\pageref{firstpage}--\pageref{lastpage}}

\maketitle

\begin{abstract}
The proposal to use \hii~galaxies (HIIGx) and giant extragalactic \hii~regions (GEHR)
as standard candles to construct the Hubble diagram at redshifts beyond the current
reach of Type Ia supernovae has gained considerable support recently with the addition
of five new HIIGx discovered by JWST. The updated sample of 231 sources now extends the
redshift range of these objects to $z\sim 7.5$, mapping the Universe's expansion over $95\%$
of its current age. In this {\it Letter} we use these sources for model selection, and show
that the $R_{\rm h}=ct$ universe is strongly favored by this probe over both 
flat-$\Lambda$CDM and $w$CDM, with relative Bayesian Information Criterion probabilities of, respectively, $91.8\%$, 
$7.4\%$ and $0.8\%$. A possible caveat with these results, however, is that 
an unknown dispersion, $\sigma_{\rm int}$, in the HIIGx standard candle relation can 
weaken the model comparisons. We find that the inclusion of $\sigma_{\rm int}$ as
an additional, optimizable parameter makes the likelihoods of flat-$\Lambda$CDM
and $R_{\rm h}=ct$ about equal, though at the expense of creating $\sim 2.5\sigma$ 
tension between our inferred matter density $\Omega_{\rm m}$ and its 
{\it Planck}-optimized value.
\end{abstract}

\begin{keywords}
\hii\ regions --- galaxies: general --- cosmological parameters --- cosmology:
observations --- cosmology: theory --- distance scale
\end{keywords}



\section{Introduction}
\hii~galaxies (HIIGx) are massive and compact, with a total luminosity almost
completely dominated by the starburst. Analogously, giant extragalactic \hii~regions
(GEHR) also contain massive bursts of star formation, though these sites tend to be
located in the outer discs of late-type galaxies. In both cases, most of the
luminosity is produced by rapidly forming stars surrounded by ionized hydrogen.
The \hbox{HIIGx} and GEHR are physically similar \citep{1987MNRAS.226..849M}, with
indistinguishable optical spectra, characterized by strong Balmer emission
lines in $\ha$ and $\hb$ due to the surrounding, ionized hydrogen
\citep{1972ApJ...173...25S,1977ApJ...211...62B,1981MNRAS.195..839T,2000A&ARv..10....1K}.

These galaxies can serve as standard candles (e.g., \citealt{2000MNRAS.311..629M,2005MNRAS.356.1117S})
because both the number of ionizing photons and the turbulent velocity of the gas 
(which is dominated by the gravity of the stars) increase as the mass of the starburst 
component rises. The luminosity in $\hb$, $L(\hb)$, is thus correlated with the ionized 
gas velocity dispersion, $\sigma$ \citep{1981MNRAS.195..839T}, with a small enough scatter 
that allows it to be used as a cosmic distance indicator (see
\citealt{1987MNRAS.226..849M,1988MNRAS.235..297M,2000AJ....120..752F,2000MNRAS.311..629M,2002MNRAS.329..481B,2003ASPC..297..143T,2005MNRAS.356.1117S,2011ApJ...735...52B,2011MNRAS.416.2981P,2012PhLB..715....9M,2012MNRAS.425L..56C,2014MNRAS.442.3565C,2015MNRAS.451.3001T}).

In their most recent analysis, \cite{2025MNRAS.538.1264C} used this correlation
to map the Hubble flow from $z\sim 0$ to $\sim 7.5$ in the context of flat-$\Lambda$CDM,
optimizing the cosmological parameters over more than 12 Gyr of cosmic expansion.
In this {\it Letter}, we continue to build support for their pioneering work by
using these same data, and the $L(\hb)$-$\sigma$ correlation, to carry out
model selection among three principal cosmologies: (i) flat-$\Lambda$CDM, (ii)
$w$CDM and (iii) the $R_{\rm h}=ct$ universe. The latter scenario has been
developed and tested using over 30 different kinds of data, demonstrating that
the observations favour it over the current standard model throughout cosmic
history, especially at high redshifts, where JWST has made several groundbreaking
discoveries (see \citealt{2023MNRAS.521L..85M,2024PDU....4601587M,2024EPJC...84.1279M,2024A&A...689A..10M}
and references cited therein). A full description of this model and many of the already 
completed tests are described in \cite{Melia2020}.

\section{Observational data and methodology}
We take the data from \cite{2025MNRAS.538.1264C}, which includes an `anchor'
sample of 36 GEHRs and nearby HIIGx with independently measured distance moduli analysed in \cite{2018MNRAS.474.1250F},
a full sample of 181 HIIGx analysed in \cite{2021MNRAS.505.1441G},
9 newer HIIGx from \cite{2023A&A...676A..53L}, and the 5 HIIGx newly
discovered by JWST \citep{2024A&A...684A..87D}, constituting a total sample
of 231 sources.

\begin{table*}
\begin{center}
\caption{Flux and gas velocity dispersion of \hii~Galaxies discovered by JWST$^{a}$.}\label{table1}
\begin{tabular}{lccc}
&&& \\
\hline\hline
Name&\emph{z}&$\log \sigma$&$\log F(\hb)$\\
    &        &$(\mathrm{km\,s^{-1}})$&$(\mathrm{erg\,s^{-1}\,cm^{-2}})$\\
\hline
&&& \\
JADES-NS--00016745 & 5.56616  $\pm$ 0.00011  & 1.731 $\pm$  0.016 &  $-17.64 \pm  0.21$ \\
JADES-NS--10016374 & 5.50411  $\pm$ 0.00007  & 1.785 $\pm$  0.014 &  $-18.14 \pm  0.20$ \\
JADES-NS--00019606 & 5.88979  $\pm$ 0.00008  & 1.622 $\pm$  0.019 &  $-18.11 \pm  0.27$ \\
JADES-NS--00022251 & 5.79912  $\pm$ 0.00007  & 1.621 $\pm$  0.011 &  $-17.86 \pm  0.11$ \\
JADES-NS--00047100 & 7.43173  $\pm$ 0.00015  & 1.868 $\pm$  0.024 &  $-17.62 \pm  0.39$ \\
&&& \\
\hline\hline
\end{tabular}
\begin{description}
  \item[$^{a}$] {Taken from \cite{2025MNRAS.538.1264C}.}
\end{description}
\end{center}
\end{table*}

The $\hb$ luminosity is calculated from the reddening corrected $\hb$ fluxes,
$F(\mathrm{H}\beta)$, and their $1\sigma$ uncertainties, as described in 
\cite{2015MNRAS.451.3001T} (see also \citealt{2000ApJ...533..682C}):
\begin{equation}\label{eq:L}
 L(\hb) = 4 \pi  D_L^2(z) F(\hb)\;,
\end{equation}
where $D_L$ is the cosmology-dependent luminosity distance at redshift $z$.
To illustrate the quality of these data, we list the corrected dispersions
and their $1\sigma$ uncertainties of the new sources discovered by JWST in
column (3) of Table~\ref{table1}, along with the measured $\hb$ fluxes and
their errors in column (4). To account for the known systematic discrepancy 
between velocity dispersion ($\sigma$) measurements derived from $\Oiii$ and 
Balmer emission lines, \cite{2025MNRAS.538.1264C} implemented a velocity 
correction of $\mathrm{2.1 \,km\,s^{-1}}$ \citep{2016MNRAS.462.2431C} for 
three JWST-NIRSpec HIIGx lacking Balmer-line $\sigma$ measurements.
This calibration factor was established through comparative analysis of
datasets containing both measurement types. For the flux corrections,
all $\hb$ measurements underwent extinction adjustment using the
\cite{2003ApJ...594..279G} dust extinction law. For JWST-NIRSpec observations,
\cite{2025MNRAS.538.1264C} determined extinction levels through the Balmer
decrement method utilizing $\ha/\hb$ flux ratios. In cases where $\ha$ fluxes
were unavailable, they adopted the mean extinction value derived from the
VUDS/VANDELS sample as a systematic baseline. The rest of the catalog may 
be found in \cite{2025MNRAS.538.1264C} and references cited therein.

The emission-line luminosity correlation with the ionized gas velocity dispersion
\citep{2012MNRAS.425L..56C,2014MNRAS.442.3565C,2015MNRAS.451.3001T} may be expressed
in the form
\begin{equation}\label{eq:L-sigma}
\log L(\hb)=\beta \log \sigma+\alpha\;,
\end{equation}
where $\beta$ is the slope and $\alpha$ is a constant representing the logarithmic
luminosity at $\log \sigma=0$. Since this luminosity indicator is cosmology-dependent,
however, the coefficients $\alpha$ and $\beta$ must be optimized simultaneously with the
rest of the cosmological parameters.

Then, with Equations~(\ref{eq:L}) and (\ref{eq:L-sigma}), we may write the distance modulus as
\begin{equation}
\mu_{\rm obs}=2.5\left[\beta \log \sigma+\alpha - \log F(\hb)\right]-100.2 \;.
\end{equation}
The corresponding error ($\sigma_{\mu_{\rm obs}}$) in $\mu_{\rm obs}$ is found via
error propagation, i.e.,
\begin{equation}\label{eq:sigma}
\sigma_{\mu_{\rm obs}}=2.5\left(\beta^{2} \sigma^{2}_{\log \sigma}+\sigma^{2}_{\log F}\right)^{1/2} \;,
\end{equation}
in terms of the $1\sigma$ uncertainties, $\sigma_{\log \sigma}$ and $\sigma_{\log F}$,
in $\log \sigma$ and $\log F(\hb)$, respectively.

By comparison, the theoretical distance modulus, $\mu_{\rm th}$, of an \hii~galaxy at redshift
$z$ is
\begin{equation}
\mu_{\rm th}\equiv5 \log\left[\frac{D_{L}(z)}{\rm Mpc}\right]+25\;.
\end{equation}
In $\Lambda$CDM, $D_L(z)$ is given as
\begin{eqnarray}\label{eq:DL_LCDM}
&\null&\hskip-0.3in D_{L}^{\Lambda {\rm CDM}}(z) = {c\over
H_{0}}\,{(1+z)\over\sqrt{\mid\Omega_{k}\mid}}\; {\rm
sinn}\Biggl\{\mid\Omega_{k}\mid^{1/2}\times\nonumber\\
&\null&\hskip-0.3in \int_{0}^{z}{dz\over\sqrt{\Omega_{\rm m}(1+z)^{3}+\Omega_{k}(1+z)^{2}+
\Omega_{\rm de}(1+z)^{3(1+w_{\rm de})}}}\Biggr\},\qquad
\end{eqnarray}
where the energy density, $\rho=\allowbreak\rho_r+\nobreak\rho_{\rm m}+\nobreak\rho_{\rm de}$,
with contributions from radiation, matter (luminous and dark), and dark energy, is
expressed as a fraction of today's critical density, $\rho_c\equiv 3c^2 H_0^2/8\pi G$,
with $\Omega_{\rm m}\equiv\rho_{\rm m}/\rho_c$, $\Omega_r\equiv\rho_r/\rho_c$,
and $\Omega_{\rm de} \equiv \rho_{\rm de}/\rho_c$. In addition, $p_{\rm de}=
w_{\rm de}\rho_{\rm de}$ is the dark-energy equation of state, and Equation~(\ref{eq:DL_LCDM})
assumes that the radiation density is negligible up to $z\sim 8$. The
spatial curvature of the Universe is represented by $\Omega_{k}=1-\Omega_{\rm m}-
\Omega_{\rm de}$---appearing as a term proportional to the spatial curvature
constant $k$ in the Friedmann equation. In Equation~(\ref{eq:DL_LCDM}) sinn is $\sinh$ when
$\Omega_{k}>0$ and $\sin$ when $\Omega_{k}<0$. In this {\it Letter}, we assume
spatial flatness throughout our analysis, for which $\Omega_{k}=0$ and the right 
side of this equation then reduces to $(1+z)c/H_{0}$ times the integral.

The luminosity distance in the $R_{\rm h}=ct$ universe
\citep{2003eisb.book.....M,2007MNRAS.382.1917M,2009IJMPD..18.1889M,2012MNRAS.419.2579M},
is given by the much simpler expression,
\begin{equation}\label{DL_Rh}
D_L^{R_{\rm h}=ct}(z)={c\over H_0}(1+z)\ln(1+z)\;.
\end{equation}

We optimize the cosmological parameters and (simultaneously) the coefficients
$\alpha$ and $\beta$ via a joint analysis involving the 195 HIIGx and the anchor
sample of 36 nearby objects, using the method of maximum likelihood estimation
(MLE; see \citealt{2015AJ....149..102W}). The final log-likelihood sampled by
the Python Markov chain Monte Carlo (MCMC) module, EMCEE \citep{2013PASP..125..306F},
is a sum of the separate likelihoods of the HIIGx and anchor samples:
\begin{equation}
\ln(\mathcal{L}_{\rm tot})=\ln(\mathcal{L}_{\rm HIIGx})+\ln(\mathcal{L}_{\rm anchor})\;,
\end{equation}
where
\begin{equation}\label{eq:likelihood1}
\mathcal{L}_{\rm HIIGx} = \prod_{i}^{195}
\frac{1}{\sqrt{2\pi}\,\epsilon_{{\rm HIIGx},i}}\;\times
\exp\left[-\,\frac{\left(\mu_{{\rm obs},i}-\mu_{\rm
      th}(z_i)\right)^{2}}{2\epsilon^{2}_{{\rm HIIGx},i}}\right]
\end{equation}
and
\begin{equation}\label{eq:likelihood2}
\mathcal{L}_{\rm anchor} = \prod_{i}^{36}
\frac{1}{\sqrt{2\pi}\,\epsilon_{{\rm anchor},i}}\;\times
\exp\left[-\,\frac{\left(\log L(\hb)_{i}-\beta \log \sigma_{i}-\alpha\right)^{2}}{2\epsilon^{2}_{{\rm anchor},i}}\right] \;.
\end{equation}
In Equation~(\ref{eq:likelihood1}), the variance on each HIIGx,
\begin{equation}\label{eq:variance1}
\epsilon^{2}_{{\rm HIIGx},i}=\sigma^{2}_{\mu_{\rm obs},i}+\left[\frac{5\sigma_{D^{\rm th}_{L},i}}{\ln{10}\,D^{\rm th}_{L}(z_i)}\right]^2,
\end{equation}
is given in terms of the propagated uncertainty, $\sigma_{\mu_{\rm obs},i}$, in $\mu_{{\rm obs},i}$
(see Equation~\ref{eq:sigma})
and the propagated uncertainty of $D^{\rm th}_{L}(z_i)$ originating from the redshift uncertainty.
In Equation~(\ref{eq:likelihood2}), the variance on each nearby object,
\begin{equation}\label{eq:variance2}
\epsilon^{2}_{{\rm anchor},i}=\sigma^{2}_{\log L,i}+\beta^{2} \sigma^{2}_{\log \sigma,i},
\end{equation}
is given in terms of the distance indicator measurement uncertainty, $\sigma_{\log L,i}$, in 
$\log L(\hb)_{i}$ and the propagated uncertainty of $\log \sigma_{i}$. Note that the first 
factor $\epsilon_{\rm HIIGx}$ in Equation~(\ref{eq:likelihood1}) (or $\epsilon_{\rm anchor}$
in Equation~\ref{eq:likelihood2}) is not a constant, since it depends on the value of $\beta$,
so maximizing $\mathcal{L}$ is not exactly equivalent to minimizing the $\chi^{2}$ statistic, 
i.e., $\chi^{2}=\sum_i\frac{\left(\mu_{{\rm obs},i} - \mu_{\rm th}(z_i)\right)^2}
{\epsilon^{2}_{{\rm HIIGx},i}}$ (or $\chi^{2}=\sum_i\frac{\left(\log L(\hb)_{i}-\beta 
\log \sigma_{i}-\alpha\right)^2}{\epsilon^{2}_{{\rm anchor},i}}$).

\section{Optimization of the model parameters}
We carry out model selection using three distinct models: flat-$\Lambda$CDM,
flat-$w$CDM (in which dark energy is not necessarily a cosmological constant),
and $R_{\rm h}=ct$.

\subsection{$\Lambda$CDM}
\begin{figure}
\vskip-0.1in
\centerline{\includegraphics[angle=0,scale=0.42]{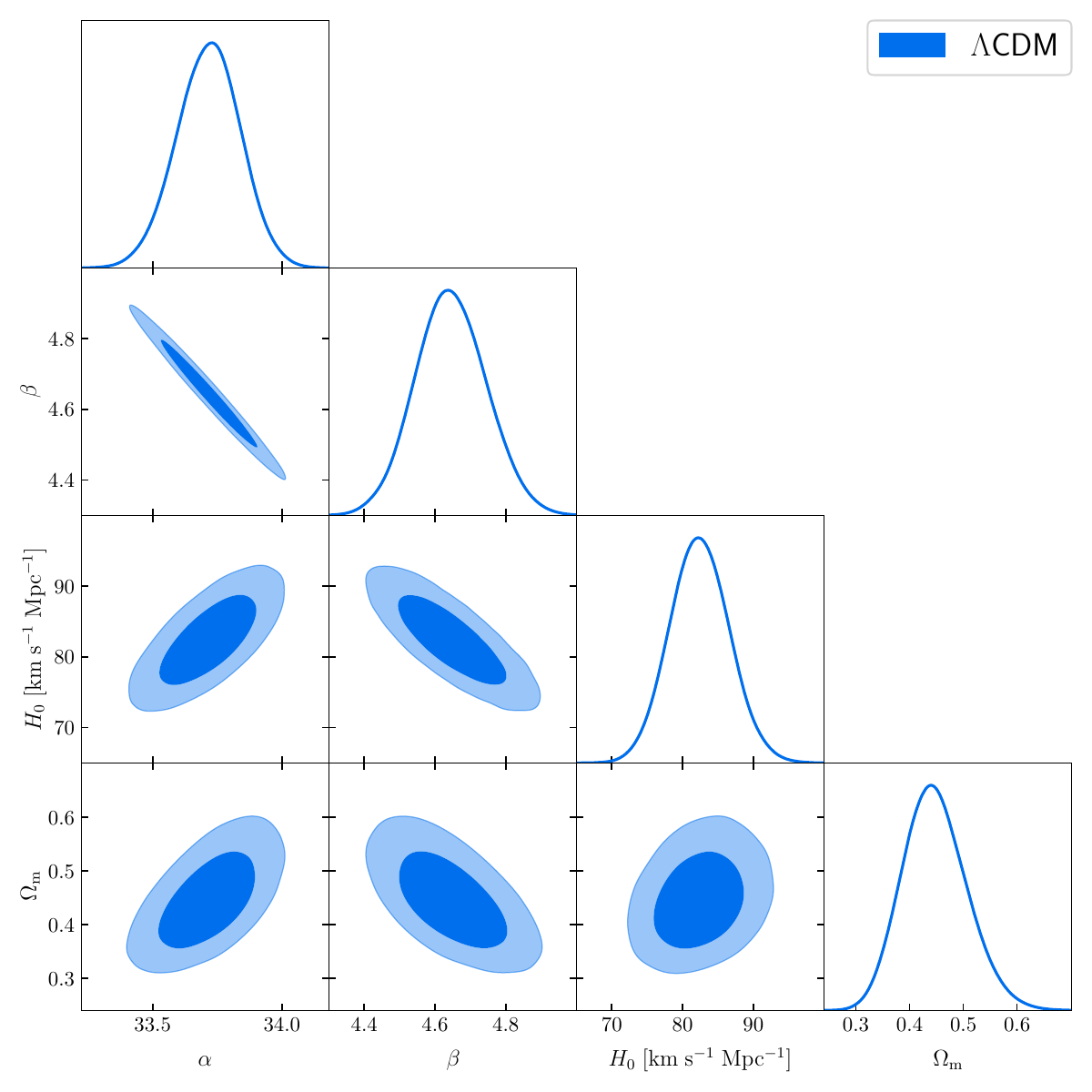}}
\caption{1-D probability distributions and 2-D regions with the $1$-$2\sigma$
contours corresponding to the parameters $\alpha$, $\beta$, $H_0$ and
$\Omega_{\rm m}$ in flat $\Lambda$CDM.}\label{LCDM}
\end{figure}

This is the most basic $\Lambda$CDM model, with $w_{\rm de}=1$.
With spatial flatness, the free parameters to be constrained are 
the two coefficients $\alpha$ and $\beta$, the Hubble constant $H_0$,
and the matter density parameter $\Omega_{\rm m}$. The 1D marginalized 
posterior distributions and 2D regions with $1$-$2\sigma$ contours 
corresponding to these four free parameters, constrained by the HIIGx 
and anchor samples, are presented in Figure~\ref{LCDM}. The optimized 
parameter values corresponding to these contours are listed in 
Table~\ref{table2}. The maximum value of the joint likelihood function
for the optimized flat $\Lambda$CDM model is given by $-2\ln \mathcal{L}=847.75$,
which we shall need when comparing models using the Bayes information criterion (BIC).

\subsection{$w$CDM}
\begin{figure}
\vskip-0.1in
\centerline{\includegraphics[angle=0,scale=0.33]{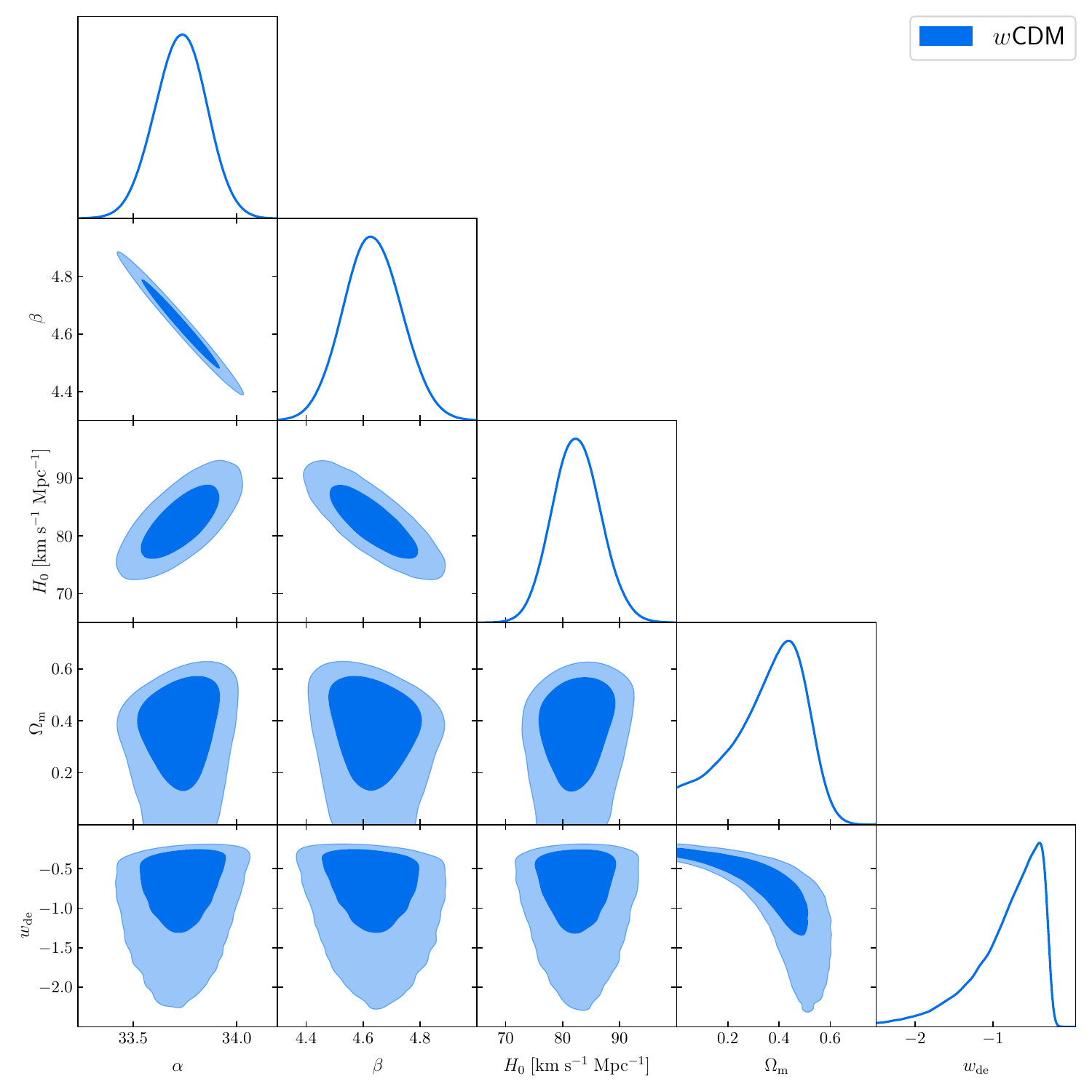}}
\caption{1-D probability distributions and 2-D regions with the $1$-$2\sigma$
contours corresponding to the parameters $\alpha$, $\beta$, $H_0$, $\Omega_{\rm m}$
and $w_{\rm de}$ in the $w$CDM model.}\label{wCDM}
\end{figure}

In the second model, we relax the assumption that dark energy is a cosmological
constant with $w_{\rm de}=-1$, and allow both $w_{\rm de}$ and $\Omega_{\rm m}$
to be free. The optimized parameters corresponding to the best-fit $w$CDM model
are shown in Figure~\ref{wCDM}, and are listed in Table~\ref{table2}. The maximum value of the joint
likelihood function for the optimized $w$CDM model is given by
$-2\ln \mathcal{L}=846.72$.

\subsection{The $R_{\rm h}=ct$ Universe}
\begin{figure}
\vskip-0.1in
\centerline{\includegraphics[angle=0,scale=0.55]{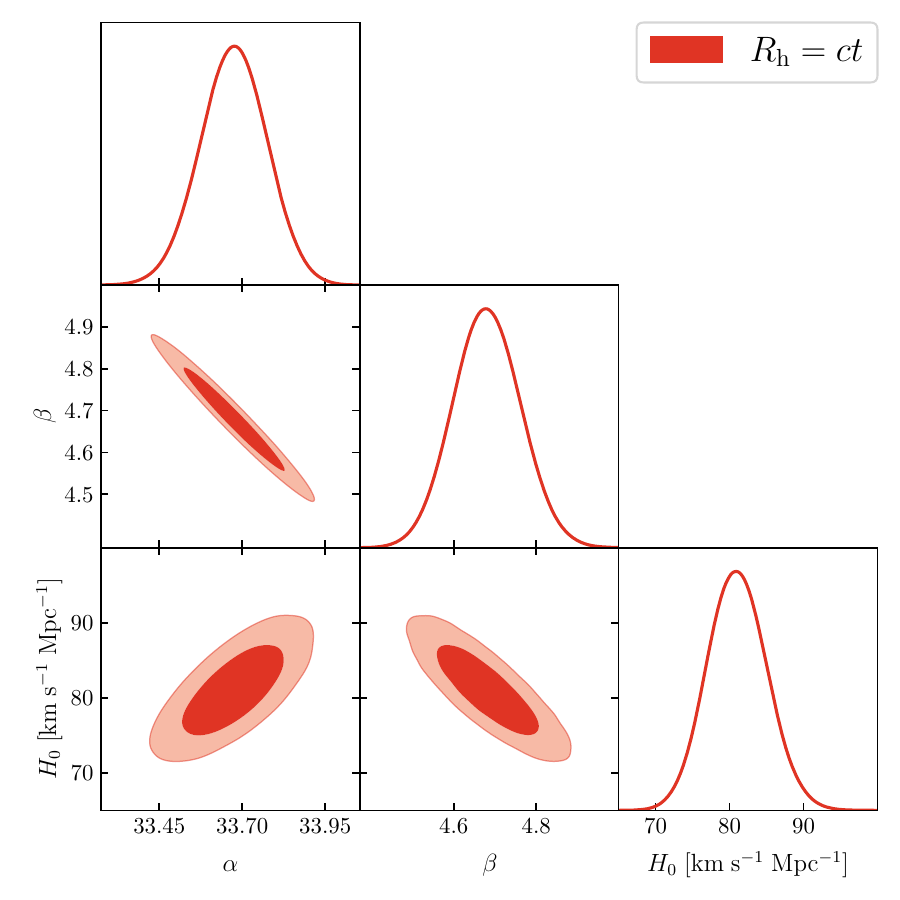}}
\caption{1-D probability distributions and 2-D regions with the $1$-$2\sigma$
contours corresponding to the parameters $\alpha$, $\beta$ and $H_0$
in $R_{\rm h}=ct$.}\label{Rhct}
\end{figure}

By comparison, the $R_{\rm h}=ct$ universe has only one free parameter, $H_{0}$.
The results of fitting the $L(\hb)$-$\sigma$ relation with this cosmology
are shown in Figure~\ref{Rhct}, and the optimized parameters are shown in Table~\ref{table2}.
The maximum value of the joint likelihood function for the optimized
$R_{\rm h}=ct$ fit corresponds to $-2\ln \mathcal{L}=848.15$.

\begin{table*}
\begin{center}
{\footnotesize
\caption{Best-fitting results in different cosmological models.}\label{table2}
\begin{tabular}{lcccccccc}
&&&&&&& \\
\hline\hline
&&&&&&& \\
Model& $\alpha$ & $\beta$ & $H_0$ & $\Omega_{\rm m}$ & $w_{\rm de}$ & $-2\ln \mathcal{L}$ &\quad BIC & Probability\\
&&&$[\mathrm{km\,s^{-1}\,Mpc^{-1}}]$ &&&& \\
\hline
&&&&&&& \\
$R_{\rm h}=ct$            & $33.67\pm 0.10$ & $4.68\pm0.08$ & $81.0\pm 4.0$ & --  & -- & 848.15 & 864.48 & $91.8\%$ \\
&&&&&&& \\
$\Lambda$CDM              & $33.72\pm0.12$ & $4.64\pm0.10$ & $82.4\pm 4.2$ &
$0.44^{+0.06}_{-0.06}$ & -- & 847.75 & 869.52 &  $7.4\%$ \\
&&&&&&& \\
$w$CDM                    & $33.73\pm 0.12$ & $4.63\pm 0.10$ & $82.4\pm 4.2$ &
$0.38^{+0.12}_{-0.18}$ & $-0.72^{+0.31}_{-0.56}$ & 846.72 & 873.93 & $0.8\%$ \\
&&&&&&& \\
\hline\hline
\end{tabular}
}
\end{center}
\end{table*}

\begin{table*}
\begin{center}
{\footnotesize
\caption{Best-fitting results in different cosmological models with 
inclusion of an intrinsic dispersion $\sigma_{\rm int}$ in the $L(\hb)-\sigma$ 
correlation.}\label{table3}
\begin{tabular}{lccccccccc}
&&&&&&& \\
\hline\hline
&&&&&&& \\
Model& $\alpha$ & $\beta$ & $\sigma_{\rm int}$ & $H_0$ & $\Omega_{\rm m}$ & $w_{\rm de}$ & $-2\ln \mathcal{L}$ &\quad BIC & Probability\\
&&&&$[\mathrm{km\,s^{-1}\,Mpc^{-1}}]$ &&&& \\
\hline
&&&&&&&& \\
$R_{\rm h}=ct$            & $34.71\pm 0.20$ & $3.83\pm0.16$ & $0.28\pm 0.02$ & $119.8^{+12.4}_{-11.3}$ & --  & -- & 496.53 & 518.30 & $48.8\%$ \\
&&&&&&&& \\
$\Lambda$CDM              & $35.04\pm0.23$ & $3.55\pm0.18$ & $0.28\pm 0.02$ & $127.6^{+13.4}_{-12.1}$ &
$0.74^{+0.15}_{-0.15}$ & -- & 491.15 & 518.36 &  $47.3\%$ \\
&&&&&&&& \\
$w$CDM                    & $35.08\pm 0.23$ & $3.52\pm 0.18$ & $0.28\pm 0.02$ & $128.3^{+13.5}_{-12.1}$ &
$0.71^{+0.19}_{-0.28}$ & $-0.72^{+0.56}_{-1.21}$ & 490.72 & 523.37 & $3.9\%$ \\
&&&&&&&& \\
\hline\hline
\end{tabular}
}
\end{center}
\end{table*}

\section{Discussion and Conclusion}
HIIGx and GEHR provide useful standard candles via the correlation of their
$\hb$ luminosity and the velocity dispersion in the gas surrounding the
ionizing stars. Unlike tracers at lower redshifts, such as Type Ia SNe,
these sources can be seen to redshifts exceeding $\sim 7.5$, as demonstrated
by the recent JWST discoveries. The pioneering work with these objects
\citep{2025MNRAS.538.1264C} has tended to focus on the optimization of cosmological
parameters in $\Lambda$CDM. As we have shown here, however, their probative power
may be even more important than this, since they can be used effectively for model
selection.

This is not to imply that JWST on its own is promoting this type of analysis,
though its contribution to the sample of HIIGx sources from the Epoch of Reionization
is allowing us to study the expansion of the Universe across over $95\%$ of its
current age. JWST has had a profound impact on our understanding of the early
Universe, producing at least four surprising results that seriously challenge
the timeline predicted by the standard model. These include: (i) the too-early
appearance of well formed galaxies at $z>14$ \citep{2023MNRAS.521L..85M}; (ii) the premature
formation of supermassive black holes at redshifts approaching $\sim 10$
\citep{2024EPJC...84.1279M}; (iii) the formation and growth of PAH grains at $z\sim 7$ that
should have taken over a Gyr to form \citep{2024PDU....4601587M}; and (iv) the reionization
crisis implied by the over-abundance of ionizing radiation at high redshifts
\citep{2024A&A...689A..10M}. It is difficult to interpret these conflicting results as mere
`tension' with the standard model. The disagreements are much more serious
than that.

This is the context within which we now consider yet another significant
anomaly uncovered by JWST, this time via the $L(\hb)-\sigma$ correlation
of giant \hii~regions. We had previously attempted to carry out model
selection using the sources from a more limited sample \citep{2016MNRAS.463.1144W},
notably from a redshift range restricted to $z< 2.3$, and concluded
even then that the model favoured by this diagnostic is $R_{\rm h}=ct$,
not $\Lambda$CDM.

With the added redshift range provided by JWST, the power of this probe
is significantly greater. And as one can see from a comparison of the
likelihoods quoted in Table~\ref{table2}, when the parameter optimization is handled
via MLE, the Bayes Information Criterion, ${\rm BIC}=-2\ln \mathcal{L}+
(\ln N)n$, where $N$ is the number of data points and $n$ the number
of free parameters \citep{1978AnSta...6..461S}, strongly
favours $R_{\rm h}=ct$ over both $\Lambda$CDM and $w$CDM. According to
a direct comparison of these three models, the former has a probability
of $\sim 92\%$ of being the correct cosmology, compared to only $\sim 7\%$
for flat-$\Lambda$CDM, and an even less favoured $\sim 1\%$ for flat-$w$CDM.
We must point out again that this outcome is based on the expansion history
of the Universe over $\sim 95\%$ of its current age.

An important caveat to this work, however, is that,
despite the clear separation in the model outcomes shown in Table~\ref{table2}, 
a possible concern with this analysis is the role played by an intrinsic scatter 
$\sigma_{\rm int}$ in the $L({\rm H}\beta)-\sigma$ correlation. This dispersion
could introduce a bias in the likelihood estimates of some cosmological models 
compared to others. To test how our conclusions may be impacted by the presence
of such a scatter, we performed a parallel comparative analysis of these 
measurements by including a global intrinsic scatter $\sigma_{\rm int}$ in 
our likelihood estimates. Since $\sigma_{\rm int}$ is not known a priori, 
we have modeled it along with the other variables in our maximum likelihood 
analysis. To do this, we have rewritten the variances in Equations~(\ref{eq:variance1}) 
and (\ref{eq:variance2}) as
\begin{equation}
\epsilon^{2}_{{\rm HIIGx},i}=6.25\left(\sigma^{2}_{\rm int}+\beta^{2} 
\sigma^{2}_{\log \sigma,i}+\sigma^{2}_{\log F,i}\right) +
\left[\frac{5\sigma_{D^{\rm th}_{L},i}}{\ln{10}\,D^{\rm th}_{L}(z_i)}\right]^2,
\end{equation}
and
\begin{equation}
\epsilon^{2}_{{\rm anchor},i}=\sigma^{2}_{\rm int}+\sigma^{2}_{\log L,i}+
\beta^{2} \sigma^{2}_{\log \sigma,i},
\end{equation}
respectively. In our MLE framework, $\sigma_{\rm int}$ is treated as an 
additional free parameter and optimized individually for each cosmological 
model (though it turns out to be identical in all three cases we consider here). 
The results are shown in Table~\ref{table3}. The likelihoods have indeed 
changed somewhat, and $\Lambda$CDM is less disfavored, though the $w$CDM 
model is still unlikely compared to the other two. The cosmological 
parameters, however, are less tightly constrained and appear to be more 
inconsistent with their {\it Planck} optimized values. In other words, 
though the inclusion of $\sigma_{\rm int}$ makes the likelihood of 
$\Lambda$CDM comparable to that of $R_{\rm h}=ct$, $\Omega_{\rm m}$ must 
deviate by over $2.5\sigma$ from the concordance model. Clearly, to 
advance the use of HIIGx in cosmology, future work must better 
characterize $\sigma_{\rm int}$ and expand the sample size.

Together with the previous four JWST conflicts with the standard model,
this result provides compelling evidence that the timeline predicted
by $\Lambda$CDM---at least in the early Universe---is not supported by
the data. Instead, all five of these discoveries point to the expansion
history predicted by $R_{\rm h}=ct$. The consequences of this conclusion
are, of course, of great interest going forward as we attempt to modify
the standard model to better account for the high-precision measurements
being made today. Further development of $R_{\rm h}=ct$, and its impact on
our interpretation of these data, will appear elsewhere \citep{Melia2026}.

\section*{Acknowledgements}
We are grateful to the anonymous referee for helpful comments.
This work is supported by the China Manned Space Program (grant No. CMS-CSST-2025-A01), 
the National Natural Science Foundation of China (grant Nos. 12422307 and 12373053), 
and the Natural Science Foundation of Jiangsu Province (grant No. BK20221562).

\section*{Data Availability}
No new data were generated or analysed in support of this research.











\bsp	
\label{lastpage}
\end{document}